
\input harvmac.tex
\def\cc{\vrule width.3cm height .4pt}
\def\dW{\Delta W}
\let\bmx=\bordermatrix
\let\bar=\overline
\def\nl{\hfill\break}
\let\<=\langle
\let\>=\rangle
\let\e=\epsilon
\let\tht=\theta
\let\t=\theta
\let\O=\Omega
\let\r=\rho
\let\l=\lambda
\let\p=\prime
\noblackbox
\pretolerance=750

\nref\rZandZ{A.B. Zamolodchikov and Al.B. Zamolodchikov, Ann. Phys. 120
(1980) 253.}
\nref\rAZi{A.B. Zamolodchikov, Adv. Stud. Pure Math. 19 (1989) 1.}
\nref\rYY{C.N. Yang and C.P. Yang, J. Math. Phys. 10 (1969) 1115.}
\nref\tbaref{Al.B. Zamolodchikov, Nucl. Phys. B342 (1990) 695.}
\nref\rMar{M.J. Martins, Phys. Rev. Lett. 67 (1991) 419.}
\nref\rKMiv{T. Klassen and E. Melzer, ``Spectral Flow between
Conformal Field Theories in 1+1 Dimensions'', Chicago preprint EFI 91-17,
Miami preprint UMTG-162.}
\nref\rPF{P. Fendley, ``Excited-state thermodynamics'', Boston preprint
BUHEP-91-16.}
\nref\zamint{A.B. Zamolodchikov, JETP Lett. 46 (1987) 160.}
\nref\rAlZiii{Al.B. Zamolodchikov, Nucl. Phys. B358 (1991) 524.}
\nref\RSOS{A.B. Zamolodchikov, Landau Institute preprint, September 1989.}
\nref\rAL{ A. LeClair, Phys. Lett. 230B (1989) 103.}
\nref\rIM{H. Itoyama and P. Moxhay, Phys. Rev. Lett. 65 (1990) 2102.}
\nref\rAlZii{Al.B. Zamolodchikov, Nucl. Phys. B358 (1991) 497.}
\nref\toda{P. Fendley, W. Lerche, S.D. Mathur and N.P. Warner,
Nucl. Phys. B348 (1991) 66;\nl P. Mathieu and M. Walton, Phys. Lett. 254B
(1991) 106.}
\nref\rVW{A.B. Zamolodchikov, Sov. J. Nucl. Phys. 44 (1986) 529;\nl
D. Kastor, E. Martinec and S. Shenker, Nucl. Phys. B316 (1989) 590;\nl
E. Martinec, Phys. Lett. 217B (1989) 431;\nl
C. Vafa and N.P. Warner, Phys. Lett 218B (1989) 51.}
\nref\rBL{D. Bernard and A. LeClair, Phys. Lett. 247B (1990) 309.}
\nref\LGSol{P. Fendley, S.D. Mathur, C. Vafa and N.P. Warner, Phys. Lett.
B243 (1990) 257.}
\nref\rABF{G.E. Andrews, R.J. Baxter and P.J. Forrester, J.Stat.Phys. 35
(1984) 193.}
\nref\rSch{K. Schoutens, Nucl. Phys. B344 (1990) 665.}
\nref\rABL{C. Ahn, D. Bernard and A. LeClair, Nucl. Phys. B336 (1990) 409}
\nref\rLW{W. Lerche and N.P. Warner, Nucl. Phys. B358 (1991) 571.}
\nref\rJR{R. Jackiw and C. Rebbi, Phys. Rev. D13 (1976) 3398.}
\nref\rFIii{P. Fendley and K. Intriligator, ``Son of
Scattering and Thermodynamics of Fractionally-Charged Supersymmetric
Solitons''.}
\nref\rWO{E. Witten and D. Olive, Phys. Lett. 78B (1978) 97.}
\nref\rGW{J. Goldstone and F. Wilczek, Phys. Rev. Lett. 47 (1981) 986}
\nref\rNS{A.J. Niemi and G.W. Semenoff, Phys. Rep. 135 (1986) 99}
\nref\rSSH{W.P. Su, J.R. Schrieffer, A.J. Heeger, Phys. Rev. Lett. 42 (1979)
1698; Phys. Rev. B22 (1980) 2099.\nl
W.P. Su and J.R. Schrieffer, Phys. Rev. Lett. 46 (1981) 738.}
\nref\dvv{R. Dijkgraaf, E. Verlinde, and H. Verlinde, Nucl. Phys. B352 (1991)
59.}
\nref\rCV{S. Cecotti and C. Vafa, ``Topological Anti-Topological Fusion''
Harvard and Trieste preprint HUTP-91/A033, SISSA-69/91/EP.}
\nref\rGep{D. Gepner, {\it Fusion Rings and Geometry} preprint
NSF-ITP-90-184.}
\nref\rLWii{W. Lerche and N. Warner, ``Solitons in Integrable $N$=2
Supersymmetric Landau-Ginzburg Models'', to appear in the Proceedings of the
Stony Brook conference ``Strings and Symmetries''.}
\nref\rRS{N. Reshetikhin and F. Smirnov, Commun. Math. Phys. 131 (1990) 157.}
\nref\rBLii{D. Bernard and A. LeClair, to appear in Commun. Math. Phys. }
\nref\rFQS{D. Friedan, Z. Qiu and S. Shenker, Phys. Lett. 151B (1985) 37;\nl
Z. Qiu, Nucl. Phys. B270 (1986) 205.}
\nref\rNR{N. Reshetikhin, J. Phys. A24 (1991) 3299.}
\nref\rPas{V. Pasquier, Nucl. Phys. B285 (1987) 162.}
\nref\rKMi{T.R. Klassen and E. Melzer, Nucl. Phys. B338 (1990) 485; Nucl. Phys
B350 (1990) 635.}
\nref\rBax{R. Baxter, {\it Exactly Solved Models in Statistical Mechanics}
Academic Press, 1982.}
\nref\rFST{E.K. Sklyanin, L.A. Takhtadzhyan and L.D. Faddeev, Theo. Math.
Phys. 40 (1980) 688. L. Faddeev, in Les Houches 1982 {\it Recent advances in
field theory and statistical mechanics} (North-Holland 1984), edited by J.-B.
Zuber and R. Stora.}
\nref\rKR{A.N. Kirillov and N. Yu. Reshetikhin, J. Phys. A20 (1987) 1587.}
\nref\rLewin{L. Lewin, {\it Polylogarithms and Associated Functions}
(North-Holland, 1981).}
\nref\rBCN{H.W.J. Bl\"ote, J.L. Cardy and M.P. Nightingale, Phys. Rev.
Lett. 56 (1986)742;\nl I. Affleck, Phys. Rev. Lett., 56 (1986) 746.}
\nref\rFG{P. Fendley and P. Ginsparg, Nucl. Phys. B324
(1989) 549.}
\nref\rKMv{T.R. Klassen and E. Melzer, ``RG Flows in the $D$ series of Minimal
CFTs'', Cornell preprint CLNS-91-1111.}
\nref\rcfun{A.B. Zamolodchikov, JETP Lett. 43 (1986) 730;\nl
A. Ludwig and J.L. Cardy, Nucl. Phys. B285 (1987) 687.}
\nref\rSW{R. Shankar and E. Witten, Nucl. Phys. B141 (1978) 349;\nl
N. Andrei and C. Destri, Nucl. Phys. B231 (1984) 445.}

\let\RSOSTBA=\rAlZii

\Title{\vbox{\baselineskip12pt\hbox{BUHEP-91-17, HUTP-91/A043}}}
{\vbox{\centerline{Scattering and Thermodynamics}
\vskip2pt\centerline{of Fractionally-Charged Supersymmetric Solitons}}}
\centerline{P. Fendley}
\bigskip\centerline{Department of Physics}
\centerline{Boston University}\centerline{590 Commonwealth Avenue}
\centerline{Boston, MA 02215}
\bigskip\centerline{K. Intriligator}
\bigskip\centerline{Lyman Laboratory of Physics}
\centerline{Harvard University}\centerline{Cambridge, MA 02138}
\vskip .3in

We show that there are solitons with fractional fermion number in integrable
$N$=2 supersymmetric models.  We obtain the soliton S-matrix for the minimal,
$N$=2 supersymmetric theory perturbed in the least relevant chiral primary
field, the $\Phi _{(1,3)}$ superfield.  The perturbed theory has a nice
Landau-Ginzburg description with a Chebyshev polynomial superpotential.  We
show that the S-matrix is a tensor product of an associated ordinary $ADE$
minimal model S-matrix with a supersymmetric part.  We calculate the
ground-state energy in these theories and in the analogous $N$=1 case and
$SU(2)$ coset models. In all cases, the ultraviolet limit is in agreement with
the conformal field theory.

\vskip .35in

\Date{11/91}


\newsec{Introduction}

Integrable models have the striking property that in a collision all momenta
are conserved individually, and that the $n$-body S-matrix factorizes into a
product of two-body ones. The enormous number of constraints this implies
means that the exact S-matrix can often be conjectured \refs{\rZandZ, \rAZi}.
The exact S-matrix encodes, nonperturbatively, physical information about our
theory; it is just a matter of extracting the information.  One way to extract
physical information from the exact S-matrix of a 1+1 dimensional integrable
theory is to do thermodynamics via the thermodynamic Bethe ansatz \refs{\rYY
,\tbaref}.  If the S-matrix is exact, the results of the TBA are
non-perturbative. For massive, integrable theories obtained from perturbing a
conformal theory by a relevant operator, for instance, we can compare the
Casimir energy predicted by the TBA with the Casimir energy as calculated in
the perturbed conformal theory.  Thus an important consistency check on a
conjectured scattering theory is to see if the Casimir energy goes over to the
correct central charge in the UV limit (where the perturbation goes away).
The next corrections to the Casimir energy can also be computed from the
conformal theory and compared with that predicted from the scattering theory.
In this way, the S-matrix contains a tremendous amount of information about
the theory.  Some excited-state energies can also be obtained in this manner
\refs{\rMar -\rPF}.

Consider, for example, perturbing the $p$-th unitary minimal (N=0) conformal
theory
by the least relevant operator, the $\Phi _{(1,3)}$ field:
\eqn\lrop{S_p\rightarrow S_p+\l \int \Phi _{(1,3)} d^2x.}
Perturbative evidence suggests that this theory has an infinite number of
conserved currents \zamint. It was explicitly shown that there is at least one
kinematic conserved current (beyond the energy-momentum tensor), which insures
that there must be completely elastic and factorizable scattering. For $\l >0$
the perturbed theory \lrop\ gives a RG trajectory flowing from the $p$-th
minimal model in the UV limit to the $(p-1)$-th minimal model in the IR limit.
Since there is a non-trivial IR limit, there must be massless particles in the
spectrum; this situation is discussed in \rAlZiii.  For $\l <0$ the perturbed
theory \lrop\ is massive and the integrability requires factorized scattering.
The exact scattering theory associated with this perturbed theory was
conjectured in \refs{\RSOS -\rIM}. The ground-state energy associated with
this conjectured scattering theory was considered using the thermodynamic
Bethe ansatz in \rAlZii\ and \rIM.  It was verified to the first few orders in
the UV limit $(\l \rightarrow 0)$ that these results agree with those obtained
{}from perturbing the conformal theory.

In this paper we will propose a scattering theory for the $N$=2 supersymmetric
analog of \lrop , i.e. the $N$=2 supersymmetric minimal models perturbed in
the least relevant operator, the $\Phi _{(1,3)}$ superfield. We assume
integrability for these theories---this has been shown to lowest order in
perturbation theory, and it is argued that the supersymmetry prevents any
corrections \toda.  These perturbed theories have a nice Landau-Ginzburg
description \rVW\ with a Chebyshev polynomial as the perturbed superpotential.

The Landau-Ginzburg analysis indicates that these $N$=2 theories are closely
related to their $N$=0 analog scattering theories \lrop .  In fact, we will
show that the S-matrix for the $N$=2 theory is a direct product of the
S-matrix for the $N$=0 analog \lrop\ with a basic $N$=2 supersymmetric
S-matrix.  This structure was conjectured in \rBL, as the simplest S-matrix
consistent with the quantum-group symmetry of the model.  Taking the
nonrenormalization of the superpotential to hold even nonperturbatively, an
assertion which has been supported in many ways, we know the exact
superpotential and the exact soliton spectrum, including the masses of the
solitons \LGSol. Thus we show that the solitons conjectured in \rBL\ do exist.
However, we have a very different interpretation of the supersymmetric part of
the S-matrix---in \rBL\ the supersymmetry is postulated to act nonlocally,
whereas here it acts locally.

The striking feature of a Chebyshev superpotential is that it has a set of
degenerate wells with equal-mass solitons interpolating between adjacent ones.
The bosonic part is the type of $N$=0 Landau-Ginzburg potential expected for
theory \lrop \RSOS \foot{This degenerate-well potential can also be predicted
by studying the associated lattice model \rABF.}. This provides a nice way of
thinking about the Landau-Ginzburg potential for the $N$=0 minimal model. It
also indicates why the soliton spectrum of the $N$=2 theory has its tensor
product structure: the bosonic $N$=0 solitons become a supermultiplet in the
supersymmetric theory.  (This tensor-product structure has also been
conjectured for the $N$=1 case \refs{\rSch, \rABL}.) We will show that this
behavior generalizes to the $N$=2 $D$-series and to the exceptional
invariants.  In fact, all the $N$=2 theories we discuss have a close relation
to an $N$=0 model. The Landau-Ginzburg soliton picture has been discussed for
many $N$=2 theories \rLW, so making this correspondence can provide intuition
concerning many non-supersymmetric models.

One fascinating feature we find is that the solitons have fractional fermion
number. This phenomena was first observed some time ago \rJR\ in a
two-dimensional $\phi^4$ theory coupled to a Dirac fermion. In the
Landau-Ginzburg picture, our models have the same structure as that treated in
\rJR. However, our models are exactly soluble and we obtain the exact
S-matrices. The models discussed in this paper have solitons with charges of
$\pm\half$. In our next paper \rFIii , we will find the S-matrix for the $N$=2
minimal models perturbed by the most-relevant operator, as well as for models
with generalized Chebyshev potentials. These theories have more
general (but still rational) fractional charges. In particular, the charges are
multiples of $1/(k+1)$ in the most-relevant perturbation of the $k$-th minimal
model.

We will provide checks on the $N$=2 scattering theories by verifying that the
Casimir energy obtained from the TBA goes over to the correct central charges
in the UV limit.  We will also perform this analysis for the analog $N$=1
discrete series \refs{\rSch,\rABL} and for the perturbed $SU(2)$ coset
models\rABL, since these fit nicely into our general picture. We will also
calculate $\tr(-1)^F$ in the $N$=2 theories as a further check on
the TBA equations.

\newsec{N=2 soliton structure}

Consider a $N$=2 theory with an effective LG description characterized by some
superpotential $W(X)$ ($X$ is a chiral superfield consisting of a complex
boson and a Dirac fermion).  Since the bosonic part of the potential is
$|W'(X)|^2$, the vacua are the points in the complex $X$ plane where $dW=0$.
The solitons $X_{ij}$ are the finite energy solutions to the equations of
motion connecting the $i$-th and $j$-th vacua: $X(\sigma =-\infty )=X^{(i)}$,
$X(\sigma = +\infty)=X^{(j)}$ (as discussed in \LGSol , not all such kinks are
to be regarded as fundamental solitons).  In the soliton sector corresponding
to a soliton $X_{ij}$ with mass $m$ and rapidity $\t$, the $N$=2 superalgebra
is \refs{\rWO ,\LGSol}:
\eqn\eSUSY{
\eqalign{&Q^2_+ = Q^2_- = \overline{Q}^2_+ = \overline{Q}^2_- =
\{ Q_+ , \overline{Q}_- \} = \{ Q_- , \overline{Q}_+ \} =0\cr
&\{ Q_+ , \overline{Q}_+ \} = 2\dW\qquad\qquad\{ Q_- , \overline{Q}_- \}
 = 2(\dW)^* \cr
&\{ Q_+ , Q_- \} = 2m e^{\theta}\qquad\qquad\{ \overline{Q}_+ , \overline{Q}_-
\} = 2m e^{-\theta} ,\cr},}
where $\Delta W=W(X^{(j)})-W(X^{(i)})$.  It follows from the above algebra
that
there is a Bogomolny mass bound $m\geq |\Delta W|$ \refs{\rWO, \LGSol}.  The
basic supermultiplet irreducible representation of \eSUSY\ for solitons
saturating this mass bound (the only type of solitons which we consider) is a
doublet consisting of solitons $u(\t )$ and $d(\t )$ with:
\eqn\SUSYrep{\eqalign{Q^-|u(\tht)\rangle &=\sqrt{2m}e^{\tht/2}|
d(\tht)\rangle\qquad\qquad \bar Q^+|u(\tht)\rangle=\omega
\sqrt{2m}e^{-\tht/2}| d(\tht)\rangle\cr Q^+|
d(\tht)\rangle&=\sqrt{2m}e^{\tht/2}|u(\tht)\rangle\qquad\qquad \bar Q^-|
d(\tht)\rangle=\omega^*\sqrt{2m}e^{-\tht/2}| u(\tht)\rangle,\cr}}
where $\omega =\Delta W/m$. All other actions annihilate the
states.

Our theory has a conserved $U(1)$ charge $F$ corresponding to fermion number.
The generators $Q^{\pm}$ have fermion number $\pm 1$, whereas $\bar Q^{\pm}$
have fermion number $\mp 1$ (this notation is less bizarre at the conformal
point where left and right fermion number are separately conserved).  In
soliton sectors the fermion number operator $F$ generally picks up an additive
constant piece, leading to fractional fermion number \refs{\rJR ,\rGW ,\rNS}.
In fact, it can be shown via adiabatic or index theorem techniques \refs{\rGW
,\rNS} that the fractional part of the fermion number in a soliton sector
$X_{ij}$ of our theory is given by
\eqn\fract{f=-{1\over 2\pi}(Im \ln W''(X))\big |^{X^{(j)}}_{X^{(i)}}.}
In our soliton doublet, $u(\t )$ has fermion number $e$ and $d(\t )$ has
fermion number $e-1$, explaining why they haven't been labeled as boson and
fermion.  We will often label the solitons by their fermion number.  The
phenomena of fractional fermion number in 1+1 dimensions occurs in physical
polymer systems where one extra bond (fermion) can be distorted into $n$
solitons and, thus, each soliton carries fermion number $1/n$ \rSSH.

The supersymmetry is defined on multi-particle states in the usual manner.
Since $Q$ is fermionic, one picks up phases when
$Q$ is brought through a particle with fermion number. For example, bringing
$Q$ through a fermion results in a minus sign. Since we will have fractional
charges, we must generalize this notion to
\eqn\Qmulti{Q^\pm|e_1 e_2\rangle= |(Q^\pm e_1) e_2\rangle +
e^{\pm i\pi e_1}|e_1 (Q^\pm e_2)\rangle,}
where the action on one soliton $(Q^\pm e)$ is defined as in \SUSYrep. The
charges $\bar Q^\mp$ act with the same phases as $Q^\pm$. In notation
analogous
to that used for coproducts in a quantum group, \Qmulti\ reads
\eqn\QQG{\Delta(Q^\pm)=Q^\pm \otimes 1 + e^{\pm i \pi F}\otimes Q^{\pm},}
where $F$ is the fermion-number operator. This similarity with the
quantum-group action is not a coincidence, since $N$=2 supersymmetry is
a special case of a quantum group \rBL.

The fractional fermion number is crucial for obtaining the correct soliton
content and S-matrices.  In \LGSol , for example, the fractional charge
structure was ignored.  It was thus mistakenly conjectured (in the context of
minimal models perturbed in the most relevant operator) on the basis of CPT
that each soliton supermultiplet should be a quadruplet---a tensor product of
two copies of the basic soliton doublet discussed above.  Taking proper
account of the fractional fermion number, one sees that this doubling is
unnecessary.  Corresponding to this doubling, the S-matrix obtained in \LGSol\
is the tensor product of the correct S-matrix with its complex conjugate.
The S-matrices for this class of integrable $N$=2 theories and the
thermodynamic calculations (which confirm the S-matrices) will be discussed in
\rFIii .  In this paper we focus on another (simpler) class of integrable $N$=2
theories, $N$=2 minimal models perturbed in the {\it least} relevant chiral
primary field.

\newsec{Chebyshev Superpotentials}

There is considerable evidence that the $N$=2 minimal models perturbed in the
least relevant chiral primary field (the $\Phi _{(1,3)}$ perturbation) are
integrable \toda.  Using the results of \dvv , the effective superpotentials
characterizing these perturbed theories are given by Chebyshev polynomials
\foot{The fact that our integrable perturbation should be in one of the
flat directions of \dvv\ is supported
by \rCV.  Our results can be viewed as additional confirmation.}:
\eqn\cheb{W_{k+2}(X=2\cos \theta )=
{2\cos (k+2)\theta\over k+2}, }
so that $W(X)={X^{k+2}/(k+2)}-X^k+\dots$. For convenience, we have set the
perturbing parameter to one (the powers of this parameter can be put back in
by charge counting; e.g., $W_5=X^5/5-
\beta X^3 +\beta ^2X$).   These perturbed theories are intimately connected
with $SU(2)_k$.  The chiral ring structure constants (in the natural basis
\dvv), for example, are the fusion rules of $SU(2)_k$ \rGep.

The vacua of our theory \cheb\ are the $k+1$ solutions of $dW(X)=\sin (k+2)\t
/\sin \t =0$:
\eqn\vax{X^{(r)} =2\cos {\pi r\over k+2}\qquad \qquad r=1,\dots ,k+1.}
Thus all the vacua are on a line in the $X$-plane. This $(k+1)$-well potential
is characteristic of the $N$=0 analog theories \lrop\ as well---the main
difference here is that we have also the fermions.  Using the methods of
\LGSol, it is easy to show that there exists a fundamental soliton connecting
each of the adjacent critical points
\vax.  Thus the spectrum consists of $k$ solitons $X_{r(r+1)}$ for $r=1,\dots
k$ and their $k$ antisolitons $X_{(r+1)r}$.  Any other possible soliton will
break apart into two or more of these solitons.  Taking our solitons to
saturate the mass bound $m\geq |\Delta W|$, using the value of the
superpotential
\cheb\ at the critical points
\eqn\watvax{W(X^{(r)})={2(-1)^r \over k+2},}
we see that our fundamental solitons connecting adjacent vacua all have equal
mass
$m=|\Delta W| =4/(k+2)$.

Each of these solitons is a supermultiplet: the doublet discussed in the last
section.  {}From \fract, we find that
$u_{r(r+1)}(\t )$ has fermion number $1/2$ and soliton $d_{r(r+1)}(\t )$ has
fermion number $-1/2$.  The corresponding $2k$ antisolitons have opposite
fermion numbers.  For the $k=1$ case this structure is just that discussed in
\rJR : the Dirac equation has one zero mode in the presence of a
soliton. (The supersymmetry requires that there be at least one; it turns out
that there is only one.) When quantized, this Dirac zero mode results in a
doublet of states, and the charge-conjugation symmetry requires that they have
charge $\pm 1/2$.

Our soliton spectrum of $4k$ particles, consisting of a
two-dimensional supermultiplet \SUSYrep\ for each soliton connecting adjacent
vacua, is exactly the structure conjectured in \rBL\ for these models;
although, as we will detail in sect.\ 4, the action of supersymmetry is
different. We will find an S-matrix for these solitons in the next section,
and in sect.\ 6 we will show that the TBA calculation gives the expected
central charges in the conformal limit.  If one ignores this supermultiplet
structure, one obtains exactly the soliton structure conjectured
\refs{\RSOS ,\rAL}
(and verified to a large extent
\rAlZii) for the $N$=0 minimal models \lrop.

\subsec{$D$-series, $E_6$, $E_7$}

The theories \cheb\ correspond to perturbing the $A_{k+1}$ minimal models in
the least relevant operator.  We found that the vacua of the theory correspond
to the nodes of the $A_{k+1}$ Dynkin diagram and the fundamental solitons
correspond to the lines in the Dynkin diagram connecting the nodes.  There are
more $N$=2 minimal models, corresponding to the $D$-series and the
exceptional models $E_6$, $E_7$ and $E_8$ \rVW. The $D$-series should be
integrable when perturbed by the least-relevant operator, because each model
is an orbifold of an $A$-series model treated in the previous subsection, and
the conserved charge discussed in \toda\ is invariant under orbifolding. One
expects that the exceptional models perturbed by their least-relevant operator
to be integrable as well. The effective Landau-Ginzburg potentials for these
perturbed theories are again obtained using the results of \dvv .  For the
$D$-series, $E_6$ and $E_7$ these potentials were obtained explicitly in
\rLWii :
\eqn\DE{\eqalign{
W_{D_k}(x,y)&={1\over k-1}\cos (k-1)\phi +{(-1)^{k-1}\over 2}xy^2
\qquad \hbox{where} \quad 2\cos \phi=2-x\cr
W_{E_6}(x,y)&={x^3\over
3}+{y^4\over 4}-xy^2+{y^2\over 2}-{x\over 12}\cr
W_{E_7}(x,y)&={x^3\over 3}+{xy^3\over 3}-x^2y+{4x^2\over 9}
-{xy\over 9}+{x\over
81}+{1\over 4374},\cr}}
where, again, we have scaled the perturbing parameter to one.  Minimizing
these potentials, one finds that the vacua of these theories correspond to the
nodes of the relevant Dynkin diagrams. We conjecture that the fundamental
solitons are the lines in the Dynkin diagrams connecting the nodes.  Just as
in the $A_{k+1}$ case, $W$ alternates in sign between connected nodes and,
thus, all solitons have equal mass.  The $N$=2 structure again requires each
soliton to consist of a $(u,d)$ doublet with charges $\pm 1/2$.
\newsec{The S-matrices}

\subsec{$X^3/3-\beta X$ (sine-Gordon at its $N$=2 point)}

The perturbed theory \cheb\ with $k=1$ corresponds to the sine-Gordon model,
with coupling at the $N$=2 point: $\beta^2={2\over 3}8\pi$ in the conventional
normalization.  The $N$=2 symmetry is a special case of the general quantum
group symmetry of the sine-Gordon model at any coupling \refs{\rRS ,\rBLii}.
Even though the $X^3/3-\beta X$ model is the ``same'' as the sine-Gordon at
this coupling, we find the action of supersymmetry in the two descriptions is
very different. The difference arises from the fact that sine-Gordon and the
$X^3/3-\beta X$ are different local projections of the same theory \rFQS. This
is analogous to the Ising model, where the spectrum can be either a free
fermion or a strongly-interacting boson with S-matrix $S=-1$. They can be
mapped on to each other by the non-local Jordan-Wigner transformation, but
they most definitely are not identical. The same type of behavior happens in
our $N$=2 model. The standard sine-Gordon S-matrix at this point \rZandZ\ does
not have an obvious supersymmetry: it must be realized nonlocally on the
sine-Gordon solitons \refs{\RSOS ,\rBL}. Even though the S-matrix in this
section is formally the same as that of sine-Gordon at the appropriate
coupling, ours does in fact allow a local action of supersymmetry. It is
likely that there is an analog of the Jordan-Wigner transformation for this
model, but we have not found one.

Fermion-number conservation means that
the S-matrix for the process $|J(\tht_1)\overline K(\tht_2)\rangle\rightarrow
|L(\tht_2) \overline M(\tht_1)\rangle $ (corresponding to a soliton $X_{12}$
colliding with a soliton $X_{21}$) is of the form
\eqn\Smat{\bmx{&u\bar u& d\bar d\cr
u\bar u&c&b\cr
d\bar d&b&c\cr}\qquad \qquad \bmx{&u\bar d &d\bar u\cr d\bar u &0&a\cr
u\bar d &a&0\cr},}
where, again, $u$ and $d$ have charges $\pm 1/2$ respectively (and their
antiparticles have opposite charges).  Demanding that the S-matrix commutes
with the supersymmetry generators with the action \SUSYrep\ and \QQG\ requires
the S-matrix elements to be
\eqn\abc{\eqalign{&a=Z(\tht)\cosh{\tht\over 2}\cr
&b=Z(\tht)i\sinh{\tht\over 2}\cr
&c=Z(\tht).\cr}}
Crossing, unitarity and the stipulation that there be no extra bound states
fixes
$Z(\tht)$ to be
\eqn\forZ{\eqalign{Z(\tht)=&{1\over\cosh{\tht\over 2}}\prod_{j=1}^{\infty}
{\Gamma^2(-{\tht\over 2\pi i}+j) \Gamma({\tht\over 2\pi i}+j+\half)
\Gamma({\tht\over 2\pi i}+j-\half)
\over \Gamma^2({\tht\over 2\pi i}+j) \Gamma(-{\tht\over 2\pi i}+j+\half)
\Gamma(-{\tht\over 2\pi i}+j-\half)}\cr
=&{1\over\cosh{\tht\over 2}}\exp({i\over 4}
\int_{-\infty}^{\infty} {dt\over t} {\sin \t \over \cosh ^2{\pi t\over
2}}). \cr}}
It is crucial in this derivation that $\Delta W/m$ is $1$ for the $X_{12}$
and $-1$ for the $X_{21}$.  Tensoring the above S-matrix with itself gives the
S-matrix obtained in \LGSol\ where, as discussed in the last section, the
basic soliton supermultiplet was unnecessarily doubled.  Our thermodynamic
calculations will confirm the above S-matrix.

This S-matrix is the same as that of the sine-Gordon model at $\beta^2={2\over
3}8\pi$ if we identify $u$ and $\bar d$ with the soliton of that
model and $d$ and $\bar u$ with the antisoliton; the fermion number in the
local description becomes the topological charge in sine-Gordon.
For purposes of the ground-state thermodynamics done in sect.\ 6,
this distinction is irrelevant; both S-matrices have the same TBA system and
hence the same Casimir energy. However, we emphasize that the interpretations
of the action of supersymmetry are very different.

\subsec{N=2 minimal models with $\Phi_{1,3}$ perturbation}

As discussed in sect.\ 3.1, the soliton structure resulting from the Chebyshev
superpotential is that of the corresponding $N$=0 minimal model with
additional $N$=2 structure. It does not immediately follow that the S-matrices
are a direct product of the $N$=0 soliton S-matrix and an $N$=2 part, because
the representation of the supersymmetry algebra depends on $\Delta W/m$, and
this can depend on which solitons are being scattered. For this reason, the
S-matrices for the $N$=2 minimal models perturbed by the most relevant
operator (to be discussed in \rFIii) cannot be a tensor product.  However,
with the $\Phi_{1,3}$ (least-relevant) perturbation, $\Delta W/m$ just
alternates between $\pm 1$ (i.e.\ $\Delta W/m=(-1)^r$ for $X_{r(r+1)}$) and,
thus, all of the two-dimensional supermultiplets obey the same supersymmetry
algebra as the $X^3-\beta X$ model discussed in the previous section. From the
point of view of the supersymmetry, every scattering process for all Chebyshev
superpotentials looks the same. Thus the S-matrix for \cheb\ is a direct
product:
\eqn\Sprod{S^{N=2}_k(\t )=S^{N=0}_{k}(\t )\otimes S^{N=2}_{k=1}(\t ),}
where $S^{N=0}_{k}$ is the S-matrix found in \refs{\RSOS ,\rAL} for the
massive scattering theory coming from the $N$=0 minimal model $SU(2)_1\otimes
SU(2)_k/SU(2)_{k+1}$ perturbed as in \lrop, and the S-matrix $S_{k=1}^{N=2}$
for the supersymmetric part is the one discussed in the previous subsection.
This direct-product structure \Sprod\ was found in \refs{\rSch,\rABL}\ for the
N=1 supersymmetric case and in \rBL\ for the $N$=2 case.  Again, the S-matrix
\Smat\ allows a local action of supersymmetry, whereas the sine-Gordon
S-matrix allows a nonlocal action. Both choices give the same
thermodynamics.  The nice thing about the LG analysis is that it gives the
Chebyshev potential and hence the soliton structure directly.

The $D_k$ series, $E_6$ and $E_7$ have the same tensor-product structure, and
the $N$=2 part is exactly the same:
\eqn\ADEprod{S_G^{N=2}(\t )=S^{N=0}_G(\t )\otimes S^{N=2}_{k=1}(\t )}
where $S^{N=2}_{k=1}$ is the basic $N$=2 S-matrix, obtained in this section,
describing how the supermultiplet elements $u$ and $d$ scatter.  The $N$=0
part $S^{N=0}_G$ is easy to find by using the correspondence between a
factorizable S-matrix and the Boltzmann weights for a lattice model. In models
with solitons, each vacuum corresponds to a ``height'' in an RSOS lattice
model; for example, the $N$=0 $A_n$ model S-matrix is equal to the Boltzmann
weights for the critical RSOS lattice model $A_{n-1}$\rABF. We conjecture that
the $D_n$ and $E$ S-matrices are the Boltzmann weights for the critical RSOS
models associated with the $D_n$ and $E$ Dynkin diagrams \rPas.

\newsec{Thermodynamic Bethe Ansatz}
The TBA allows us to extract the Casimir energy for our theory on a circle of
length $R$.  The only required input are the soliton masses and their two-body
elastic S-matrix.  The basic idea is to compute the minimum free energy of a
system of solitons at temperature $T\equiv 1/R$ living on a circle of length
$L\rightarrow\infty$.  This is accomplished by finding the allowed energy
levels on the circle and then filling the levels so as to minimize the free
energy $F$.  The TBA analysis is thus in the grand canonical picture, with a
chemical potential $\mu$ \refs{\tbaref,\rKMi}
 (general chemical potentials enter in the calculation
of excited-state energies\rPF ).  Our system can be viewed as a
two-dimensional system on a torus with length $L$ corresponding to the
compactified space and length $R=1/T$ corresponding to compactified time.  It
thus follows that the ground-state energy $E(R)$ of our system on a circle of
periodicity $R$ is given in terms of our minimum free energy by
\eqn\ground{E(R)=RF_{min}(R,L)/L \qquad \hbox{as} \quad L\rightarrow \infty ,}
when $\mu=0$.

Consider $N$ particles with rapidities $\t _1,\dots ,\t _N$ on the circle.
The allowed wavefunction must be invariant upon bringing every rapidity $\t
_k$ around the circle, through the others and back to its starting point.  In
other words we must have
\eqn\trip{e^{im_k\sinh \t _k L}T(\t _k |\t _{k+1},\dots \t_N,\t_1,\dots,
\t _{k-1})\psi =\psi,}
where $T(\t _k|\t _{k+1},\dots ,\t _{k-1})$ is the transfer matrix for
bringing particles with rapidity $\t _k$ through the others.  Our allowed
wavefunctions must satisfy \trip\ for every allowed rapidity.  In other words,
the allowed wavefunctions must be simultaneous eigenvectors of the transfer
matrices for bringing each rapidity through the others and the eigenvalues
must satisfy \trip\ (the Yang-Baxter relation ensures that the transfer
matrices commute).  This condition quantizes the allowed particle energies
$E_k=m_k\cosh \t _k$ on the circle.  We thus need to know how to diagonalize
the transfer matrix in order to even write down the TBA system of equations.
It is for this reason that the TBA equations have been derived for only a few
theories with non-diagonal S-matrices \refs{\rAlZii ,\rNR}. In the next
subsection we will review the warmup case where the S-matrix is diagonal.  In
the following subsection we will discuss some aspects of our case where the
S-matrix is not diagonal.

\subsec{The TBA equations for diagonal S-matrices}

When the S-matrix is diagonal, i.e. the only scattering is of the type $a(\t
_1)b(\t _2)\rightarrow S_{ab}(\t _1-\t _2)b(\t _2)a(\t _1)$, the transfer
matrix in \trip\ is diagonal for the $N$ particle wavefunction describing
particles of
rapidity $\t _i$ with the $i$-th particle having definite species $a_i$.  The
single-valuedness condition \trip\ then becomes simply:
\eqn\quant{e^{im_k\sinh \t _kL}\prod_{j\ne k} S_{a_ka_j}(\theta_k-\theta_j) =
1,}
an interacting-model generalization of the one particle relation $p_k=2\pi
n_k/L$.  In the large $L$ limit \quant\ gives the distribution $P_a(\t )$ of
allowed rapidity levels for particles of species $a$ in terms of the
distributions $\rho _b(\tht')$ of rapidity levels actually occupied by
particles of species $b$:
\eqn\forrho{2\pi P_a(\tht)= mL\cosh(\theta)+
\sum_b \int d\tht'\rho_b(\tht')\phi_{ab}(\theta-\theta'),}
where $$\phi_{ab}(\tht)\equiv -i{\del\ln{S_{ab}(\tht)}\over\del\tht}.$$
It is convenient
to define quantities $\e _a(\t )$ by
\eqn\pe{{\r _a(\t )\over P _a(\t )}={\l_a e^{-\e _a(\t )}\over 1+\l_a
e^{-\e _a(\t)}},}
where the fugacity $\l_a=e^{\mu_a R}$.
This definition reflects the fact that all particle species in this paper
(bosons and fermions) have $S_{aa}(0)=-1$ and, thus, at most one particle of a
given species occupies a given momentum level \foot{We define the S-matrix
elements so that free fermions scattering off of one another have S=$-1$: this
corresponds to ordering the creation operators by a particle's position in
space. This definition enables us to use the Yang-Baxter equation and do the
thermodynamics ignoring the fact that there are particles with non-zero
fermion number.}.

The distributions $\r _a (\t )$ are now chosen so as to minimize the free
energy.  This yields the TBA integral equations for the $\e _a(\t )$
in the presence of a general chemical potential \refs{\tbaref,\rKMi}:
\eqn\TBA{\e _a(\t )=m_aR\cosh (\t )-\sum _b \int {d\t ^{\p}\over 2\pi}\phi
_{ab}(\t -\t ^{\p})\ln(1+\l_b e^{-\e _b(\t ^{\p})}).}
Using \ground\ it can then be shown that the ground-state energy of the system
can be written in terms of the $\e _a$ as:
\eqn\fe{E_{\l}(R)=-\sum _a {m_a\over 2\pi }\int d\t
\cosh \t \ln (1+\l_a e^{-\e _a(\t)}).}

Having obtained the system \TBA\ appropriate for a given scattering theory,
one can compute the Casimir energy $E(R)$ by (numerically) solving this set of
coupled integral equations.  The UV limit $(m\rightarrow 0)$ and the IR limit
($m\rightarrow\infty$) of the Casimir energy can be obtained analytically and
are discussed in sects.\ 5.3 and 5.4 respectively.

\subsec{Obtaining TBA Equations for non-diagonal S-matrices}

We will describe the transfer matrices in \trip\ (one matrix for each particle
$k$) in terms of a single transfer matrix \foot{We are grateful to N.
Reshetikhin for this presentation.}.  Denote the transfer matrix for bringing
a particle of type $a$ and rapidity $\t$ through $N$ particles and ending up
with a particle of type $b$ by $T_{ab}(\t )$.  We are here suppressing the
dependence on the rapidities $\t _1,\dots \t _N$ of the $N$ particles as well
as the $q^N$ labels $c_i$ for the species of the $N$ particles before the
scattering process and the $q^N$ labels $d_i$ for the $N$ particles after the
process ($q$ is the number of different particle species).  The components of
$T_{ab}(\t )$ can be written in terms of the S-matrix elements as
\eqn\Tab{(T_{ab}(\t))_{c_i}^{d_i}\equiv
\sum_{\hbox{all} k} S_{ac_{1}}^{d_{1}k_{1}}(\tht -\tht_{1})
S_{k_{1}c_{2}}^{d_{2}k_{2}}(\tht-\tht_{2})\dots
S_{k_{N-1}c_{N}}^{d_{N}b}(\tht-\tht_{N})}
Pictorially, this is much easier to digest; it is
\bigskip
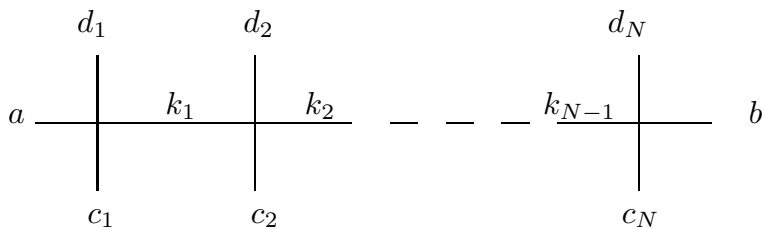
\hbox{$\hskip2cm$
\rlap{\raise.85cm\hbox{$\hskip2.4cm k_1\hskip 1.2cm k_2\hskip 2.3cm k_{N-1}$}}
\rlap{\raise50pt\hbox{$\hskip1.3cm d_1 \hskip1.5cm d_2\hskip3.7cm
d_N$}}
\rlap{\lower10pt\hbox{$\hskip1.3cm c_1 \hskip1.5cm c_2\hskip3.8cm
c_N$}}
$\hskip.3cm$\rlap{\raise.75cm\hbox{$a$ {\vrule width3.5cm height .4pt}
$\hskip.2cm$ \cc $\hskip.2cm$ \cc $\hskip.2cm$ \cc $\hskip.2cm$
{\vrule width1.7cm height .4pt} $\hskip.3cm b$}}
$\hskip.75cm$ \hbox{\vrule width.4pt height1.5cm} $\hskip1.5cm$
\hbox{\vrule width.4pt height1.5cm} $\hskip4cm$
\hbox{\vrule width.4pt height1.5cm}}

\bigskip
\noindent
where each intersection is an S-matrix element.  Using the fact that all of
our S-matrices satisfy $S_{ab}^{cd}(0)=-\delta_{ac}\delta_{bd}$,
scatterings at zero relative rapidity permute the colliding particles (with a
sign), it is seen that the matrix $\tr _oT(\t )\equiv \sum _{a}T_{aa}(\t )$
satisfies
\eqn\NRtrans{-\tr _oT(\t =\t _k)=T(\t _k|\t _{k+1},\dots ,\t _{k-1}),}
the transfer matrices appearing in \trip .  We can thus concentrate on
diagonalizing $\tr _oT(\t)$ for general $\t$ and then set $\t$ to the
different $\t _k$ at the end.  The Yang-Baxter relation ensures that the
$\tr_o T(\t )$ commute for different $\t$ and thus can be simultaneously
diagonalized for all $\t$ by a $\t$-independent set of eigenvectors; only the
eigenvalues depend on $\t$.
%

We do not need to find the eigenvalues of $\tr _o T(\t)$ explicitly in order
to do the thermodynamics; rather, we derive constraint equations so that we
can define eigenvalue densities just like the rapidity densities.  This
problem is equivalent to diagonalizing a transfer matrix in an integrable
lattice model, where the S-matrix elements play the role of the lattice
Boltzmann weights.  There are a variety of techniques to do this \rBax. One
which is particularly elegant and which is applicable to our models is the
Algebraic Bethe Ansatz \rFST. We will discuss this in appendix A.  In the next
section we apply these constraint equations to a variety of models. The result
is that we will be able to obtain a TBA system of equations of the
form \TBA \fe\ containing some extra fictitious, zero-mass, particle species to
account for the eigenvalue contribution to
\trip .

\subsec{Casimir Energy in the UV Limit}
The free energy \fe\ can be found explicitly in the $m_aR\rightarrow 0$
limit \refs{\rKR,\tbaref}. For zero chemical potential, it is given by
\eqn\uvlim{E(m_aR\rightarrow 0)\sim -{1\over \pi R} \sum _a [{\cal
L}({x_a\over
1+x_a})-{\cal L}({y_a\over 1+y_a})],}
where ${\cal L}(x)$ is Rogers dilogarithm function\rLewin
$${\cal L}(x)=-{1\over 2}\int_0^x dy\left[{\ln y\over (1-y)} + {\ln(1-y)\over
y}\right],$$
$x_a=\exp(-\epsilon(0))$, and $y_a=\exp(-\epsilon(\infty))$. The ground-state
energy is found by setting all $\l_a=1$.  It follows from
\TBA\ that the constants $x_a$ are the solutions to the equations
\eqn\xa{x_a=\prod _b (1+\l_b x_b)^{N_{ab}},}
where $N_{ab}={1\over 2\pi}\int d\t \phi _{ab}(\t )$.  (We restore the
chemical potentials because we will use them in sect.\ 6.3)
The constants $y_a$ in
\uvlim\ are nonzero only for those species $a ^{\p}$ with $m_{a ^{\p}}=0$,
where they are the solutions to
\eqn\ya{y_{a^{\p}}=\prod _{b^{\p}} (1+\l_b y_{b^{\p}})^{N_{a^{\p}b^{\p}}},}
where $b^{\p}$ also runs only over massless species.  The UV limit Casimir
energy \uvlim\ is to be compared with that of the conformal theory
\eqn\central{E(R)={2\pi  \over R}\big( h+\bar h -{c\over 12}\big),}
where $(h, \bar h)$ are the conformal dimensions of the operator which create
this state.  For unitary theories, the ground state has $h=\bar h=0$ in
\central .  We can check a proposed scattering theory by comparing \uvlim\
with \central .  Higher order in $m_aR$ corrections to
\uvlim\ can be numerically computed from \TBA\ and compared to perturbation
theory about the UV limit conformal theory.

\subsec{Casimir Energy in the IR limit}
The Casimir energy in the IR limit can be expanded in a series
in $e^{-m_aR}$.  From \TBA\ it is easy to show that for $m_aR$ large,
\eqn\TBAIR{\e_a(\tht)\approx m_aR\cosh\tht - \sum_{b'} N_{ab'}
\ln(1+\l_{b'}y_{b'})}
Plugging this into \fe, one finds
\eqn\feIR{E(mR\rightarrow\infty)\rightarrow -\sum_a q_a\int d\tht {m_a\over
2\pi}\cosh\tht e^{-m_a R\cosh\tht},}
where $$q_a=\prod_{b'}(1+\l_{b'}y_{b'})^{N_{ab'}}.$$
This provides a useful check on the results: $q_a$ must be the number of
particles of mass $m_a$ because in the $R\rightarrow\infty$ limit we have a
very dilute gas of particles, and the first term in this expansion of the free
energy is given by one-particle contributions only. In the case where there
are solitons, $q_a$ is not required to be an integer: it is given by
$B^{1/N}$, where $B$ is the number of $N$-particle configurations.

\newsec{TBA for Minimal Models Perturbed in the Least Relevant Operator}

In this section we obtain and analyze the TBA system of equations for the
$N$=2 scattering theories discussed in sect.\ 4.  We will also consider the
analogous $N$=1 case and the case of all perturbed $SU(2)_l\otimes
SU(2)_k/SU(2)_{k+l}$ coset models, since this involves little additional work.
{}From the form of the S-matrix \Sprod\ we see that the $N$=0 case enters as an
ingredient for our $N$=2 case (and also for the $N$=1 case
\refs{\rSch,\rABL}).  The RSOS S-matrix conjectured in \RSOS\ to describe the
$N$=0 minimal models $SU(2)_1\otimes SU(2)_k/SU(2)_{k+1}$ perturbed in the
least relevant operator is believed to lead to a TBA system \TBA\ which can be
described by the figure \RSOSTBA

\bigskip
\centerline{\hbox{\rlap{\raise15pt\hbox{$\ \  1\hskip.87cm 2\hskip.87cm
3\hskip1.5cm k-1\hskip.6cm k$}}
$\bigcirc$------$\bigcirc$------$\bigcirc$-- -- --
--  -- --$\bigcirc$------{\raise1pt\hbox{$\bigotimes$}}}}
\bigskip
\noindent
Each node in the figure represents a particle species for the equations \TBA ;
the $k-1$ open nodes have $m_a=0$ while the node labeled $\otimes$ is massive.
The elements $\phi _{ab}$ in \TBA\ are zero unless the nodes associated with
species $a$ and $b$ are connected by a line, in which case $\phi _{ab}(\t
)=(\cosh (\t ))^{-1}$.  In other words, $\phi _{ab}(\t )=l_{ab}\phi (\t)$
where $l_{ab}$ is the incidence matrix for the above diagram.  This TBA system
of coupled integral equations was analyzed in \RSOSTBA; the system was proven
to describe the RSOS S-matrix only for $k=1$ and $k=2$, but there is much
evidence that the correspondence holds for the whole series. We shall assume
that this is true in this section.

It is straightforward to take the UV and IR limits in the manner of sects.\ 5.3
and 5.4. With this $\phi_{ab}$, we have $N_{ab}=\half l_{ab}$. The solution to
the equations \xa\ is \refs{\rKMi,\RSOSTBA}
\eqn\anxa{x_a=\left({\sin{\pi(a+1)\over k+3}\over\sin{\pi\over k+3}}\right)^2
-1.}
To solve the equations \ya, we notice that this case corresponds to the above
Dynkin diagram with the $k$-th node removed. Thus the solution is
\eqn\anya{y_a=\left({\sin{\pi(a+1)\over k+2}\over\sin{\pi\over k+2}}\right)^2
-1.}
Plugging these into \uvlim, and using the amazing dilogarithm identity
\refs{\rKR,\rLewin}
\eqn\amazing{\sum_{a=1}^k {\cal L}\left({\sin^2{\pi\over k+3}
\over\sin^2{\pi(a+1)\over k+3}}\right)={2k\over k+3} {\pi^2\over 6}}
along with ${\cal L}(1-x)={\cal L}(1)-{\cal L}(x)$ and ${\cal L}(1)=1$,
yields
\eqn\anE{E(mR\rightarrow 0)\rightarrow
-\left({\pi\over 6R}\right)(1-{6\over{(k+2)(k+3)}}).}
This, of course, is the correct UV limit. The first corrections to \anE\ also
agree with the predictions from perturbed conformal field
theory \refs{\RSOSTBA,\rKMiv}.  The infrared limit gives a nice check: the
number of $N$-particle configurations for large $N$ for the $k$-well soliton
structure is given by $q^N$, where $q=2\cos{\pi\over k+2}$ is the largest
eigenvalue of the incidence matrix $l_{ab}$. Using the $y_a$ in
\feIR\ gives precisely this result.

\subsec{$N$=2 $X^3-\beta X$}

In this section, we apply the discussion of the appendix to obtain the TBA
equations for our basic $N$=2 S-matrix $S^{N=2}_{k=1}$ \abc .  The eigenvalues
of the $N$-particle transfer matrix are obtained from (A.10) to be
\eqn\Tmatevs{\l (\t ;\{ y\})={\cal A}\prod _{j=1}^NZ(\t -\t _j)\prod
_{r=1}^n\cosh({\t -y_r\over 2})\prod _{r=n+1}^Ni\sinh({\t -y_r\over 2}),}
where $Z(\t)$ is given by \forZ\ and where we define
\eqn\Aconst{{\cal A}\equiv \prod _{r=1}^N (i\sinh({\t -y_r\over 2}))^{-1}(
\prod _{j=1}^Ni\sinh({\t -\t _j\over 2})+(-1)^n\prod _{j=1}^N\cosh({\t -\t
_j\over 2})).}
The $y_r$ in \Tmatevs\ are the $N$ distinct solutions to (A.9), which here
becomes
\eqn\yc{\prod _{j=1}^N{i\sinh ({y_r-\t _j\over 2})\over \cosh
({y_r-\t _j\over 2})}=(-1)^{n+1},}
for any choice of $n=1, \dots , N$.  Note that, given \yc , $\cal A$ has no
poles in $\t$ and is bounded at infinity and, thus, does not depend on $\t$.

Unitarity of the S-matrix means that the eigenvalues \Tmatevs\ have magnitude
one, which is required for \trip.  The corresponding solutions to condition
\yc\ are of the form $y=z_0+i\pi/2$ and $y=z_{\bar 0}-i\pi/2$, with $z_0$ and
$z_{\bar 0}$ real.  Defining the distributions of such solutions as
$P_{0}(z_0)$ and $P_{\bar 0}(z_{\bar 0})$ respectively, the log of \yc\ gives
\eqn\Ponetwo{2\pi P_i(z_i)=\int d\t {\rho (\t )\over \cosh (z_i-\t
)},}
where $i=0,\bar 0$.

Condition \trip\ yields the relation
\eqn\allow{mL\cosh \t +Im{d\over d\t }\ln \l (\t )=2\pi P(\t )}
for the distribution of allowed rapidities. Thus the level density for the
real particles depends on the eigenvalue distributions, just as \forrho\
depends on the rapidity distribution for all the particles.  Each of the $2^N$
eigenvalues $\l(\tht,\{ y\})$ is determined by specifying whether a given
$y_r$ contributes a $\cosh(\half(\tht-y_r))$ or an $i\sinh(\half(\tht-y_r))$
to the product in \Tmatevs. (Remember that a given $y_r$ appears exactly once
in this product.) We thus define the density $P^-_i(z_i)$ of $z_i$ which
contribute a $\cosh$ to the product ($n$ contributions in total), and
$P^+_i(z_i)$ for the remaining contributions.  We thus have the relation
$P_i(z_i)$=$P^-_i(z_i)$+$P^+_i(z_i)$. With these definitions, \Tmatevs\
becomes
\eqn\Imdlog{
\eqalign{Im {d\over d\t}\ln \l (\t )&=\int d\t '\rho (\t ')Im
{d\over d\t }\ln Z(\t -\t ')\cr  &+{1\over 2} \sum
_{i=1}^2\int dz_i{P^+_i(z_i)-P^-_i(z_i)\over \cosh (z_i-\t )},\cr}}
Now from \forZ , \Ponetwo , and the integrals
$${1\over 4}\int dt {\cos \t t \over \cosh ^2({\pi t\over 2})}=\int {dy\over
2\pi }{1\over \cosh (y-\t )\cosh y}={\t \over \pi \sinh \t },$$
we obtain
$$\int d\t '\rho (\t ')Im {d\over d\t}\ln Z(\t -\t ')={1\over 2}\sum
_{i=1}^2\int dz_i{P_i(z_i)\over \cosh (z_i-\t )}.$$
Relations \allow\ and \Imdlog\ then become
\eqn\finaltrip{mL\cosh \t +\sum _{i=1}^2\int dz_i{P_i^+(z_i)\over \cosh (z_i-
\t )}=2\pi P(\t ).}

The relations \Ponetwo\ and \finaltrip\ are constraints on the particle
densities just like \forrho. Thus we can proceed identically to the diagonal
case. We minimize the free energy with respect to $\rho$ and $P_i^+$, subject
to the constraints \Ponetwo\ and \finaltrip. Defining
\eqn\pseudos{{\rho (\t)\over P (\t )}={e^{-\e (\t)}\over  1+e^{-\e (\t
)}}\quad {P^+_i(z_i)\over P_i(z_i)}={e^{-\e _i (z_i)}\over 1+e^{-\e _i
(z_i)}}}
for $i=0,\bar 0$, it is seen that we obtain a TBA system
system of the form \TBA\ and \fe\ consisting of three species, one massive and
two massless, with $\phi _{ab}(\t )=(\cosh \t )^{-1}l_{ab}$, where $l_{ab}$ is
the incidence matrix for

\bigskip
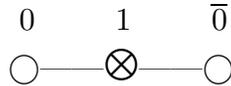
\centerline{\hbox{\rlap{\raise15pt\hbox{$\ \,\, 0\hskip.89cm 1\hskip.89cm
\bar 0$}}}
$\bigcirc$------{\raise1.5pt\hbox{$\bigotimes$}}------$\bigcirc$}
\bigskip
\noindent
The open nodes in the figure correspond to the massless pseudoparticles (the
$P_i^+$ densities) and the solid node corresponds to the soliton ($\rho$
density) with mass $m=|\Delta W|$.

To take the ultraviolet limit of this TBA system, we use our analysis of the
non-supersymmetric case. Since $l_{ab}$ is the incidence matrix for $A_3$, we
can read off the $x_a$ from \anxa: $x_0=x_{\bar 0}=2$ and $x_1=3$. The $y_a$
can be found by removing the center node, leaving two unconnected nodes and
$y_0=y_{\bar 0}=1$. Using the dilogarithm identity
\amazing\ yields
\eqn\athreeE{E(mR\rightarrow 0)\rightarrow
-\left({\pi\over 6R}\right).}
Thus our TBA system has the correct central charge $c=1$ in the UV limit.
The infrared limit also gives the correct result: we have $2^N$ different
$N$-particle configurations.

\subsec{$N$=2 minimal model with $\Phi _{(1,3)}$ perturbation}

To find the TBA system for the entire $N$=2 series, we must diagonalize the
transfer matrix \Tab\ associated with the S-matrix \Sprod. Since the S-matrix
is a tensor product of an $N$=0 S-matrix with a supersymmetric one, we can
diagonalize the two parts separately.\foot{ We emphasize that the model is not
a product of two models, since that would yield a direct product of
S-matrices.  In our situation, a particle can be thought
of as a bound-state of a particle in the $k$-th minimal model with one in the
$N$=2 $X^3-X$ model.} Each diagonalization has already been done: the $N$=0
minimal-model part was discussed at the beginning of this section, while the
supersymmetric part was done in the last subsection. Each piece contributes
their associated pseudoparticles (the open nodes in the diagrams); the
tensor-product structure means that the pseudoparticles from each piece do not
couple to ones from the other. However, both types of pseudoparticle couple to
the same massive particle, because both contribute to the single equation
\allow. Therefore, the TBA system has a single massive particle. It follows
that the TBA equations are described by the $N$=0 figure for $S^{N=0}_k$
joined at the massive node to the basic $N$=2 part:
\bigskip

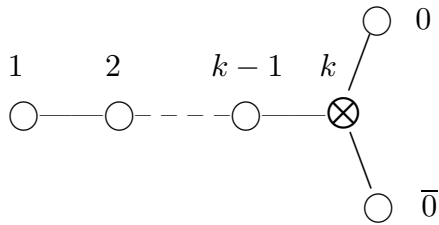
\centerline{\hbox{\rlap{\raise30pt\hbox{$\hskip4.5cm\bigcirc\hskip.25cm 0$}}
\rlap{\lower29pt\hbox{$\hskip4.4cm\bigcirc\hskip.3cm \bar 0$}}
\rlap{\raise17pt\hbox{$\hskip4.1cm\Big/$}}
\rlap{\lower14pt\hbox{$\hskip4.0cm\Big\backslash$}}
\rlap{\raise15pt\hbox{$\ 1\hskip.9cm 2\hskip1.0cm k-1\hskip.4cm k$}}
$\bigcirc$------$\bigcirc$-- -- --
--$\bigcirc$------{\raise1pt\hbox{$\bigotimes$}} }}

\bigskip

\noindent
We do not know of any significance to the fact that this is a
$D_{k+2}$ Dynkin diagram.

We can now compute the UV limit of the Casimir energy.  The $x_a$ are given by
\rKMi
\eqn\xadk{\eqalign{x_0=x_{\bar 0}&=k+1\cr
 x_a&=(a+1)^2-1\cr}}
To find the $y_a$, we remove the massive node, leaving us with the $A_{k-1}$
Dynkin diagram plus two individual nodes. Thus the $y_a$ for $a=1\dots k$ are
given by \anya, while $y_0=y_{\bar 0}=1$. There exists another amazing
dilogarithm identity \rKR\ (used in the $D_n$ section of \rKMi ) which gives
us
\eqn\anE{E(mR\rightarrow 0)\rightarrow
-\left({\pi\over 6R}\right){3k\over k+2},}
in agreement with the central charge ${3k\over k+2}$ for the $N$=2 minimal
series. The IR limit is correct as well.  It can also be shown from the TBA
analysis that the bulk contribution to $E(R)$ (terms proportional to e.g. $R$
or
$R\ln R$ in the small $R$ expansion) vanishes.  This is to be expected
since our theories are supersymmetric.  We will now calculate $\tr (-1)^F$
as another check on the supersymmetric nature of our theories.

\subsec{$\tr(-1)^F$}

To calculate $\tr(-1)^F$, we use the fact that with chemical potentials
$\mu_a$, the free energy is equal to $-T \ln \tr(e^{\mu_a N_a/T})$. Thus in a
theory with a diagonal S-matrix, one can calculate $\tr(-1)^F$ by setting the
chemical potentials $\mu_a=i\pi e_a T$, where $e_a$ is the fermion number of a
particle of species $a$. Finding the appropriate chemical potentials in a
non-diagonal scattering theory is more subtle, because the diagonalization of
the transfer matrix means that we are not working in a basis of particle
eigenstates. However, the thermodynamic expectation value of a symmetry
operator is well-defined on each state in this new basis because these
operators commute with the Hamiltonian and hence the transfer matrix.
Therefore, we need to find the eigenvalue of the symmetry operator associated
with each eigenvalue of the transfer matrix.  With some methods of determining
transfer matrix eigenvalues, it is not obvious how to do this. For example, in
the inversion-relation method used in \rAlZii, one finds the constraint
equations for the eigenvalues without constructing the eigenstate. However,
using the Algebraic Bethe Ansatz, we have an explicit expression (A.5) for the
eigenstates, making this identification simple.

In our series of models, the particles $(u,d)$ have fermion
number ($\half,-\half$).  Our eigenvector $\psi$ (A.5) thus has fermion
number $n-(N/2)$.  This is the fermion number associated with each eigenvalue
\Tmatevs. By definition, we have
\eqn\nandN{\eqalign{\int P_0^+ + \int P_{\bar 0}^+ &=N-n\cr
\int \rho &= N.\cr}}
To calculate $\tr(-1)^F$, we set $\mu_0 R=\mu_{\bar 0}R=-i\pi$, and
$\mu_k R={i\pi\over 2}$. The remaining chemical potentials for the
pseudoparticles $1,\dots,(k-1)$ remain zero; these particles arise from
diagonalizing the part of the S-matrix independent of fermion number.

To finish the calculation, we need to calculate the $x_a$ and $y_a$ from
\xa\ and \ya. The result is that with these chemical potentials, $x_a=y_a$.
Specifically, $x_k=y_k=0$, $x_0=y_0=x_{\bar 0}=y_{\bar 0}=1$, and the
remaining ones are as in \anya. The fact that $x_a=y_a$ means that $E_\l
(R)=0$ for all $R$---not only in the UV limit. Thus
\eqn\wit{\tr(-1)^F=e^{-E(R)L}=0 + {\cal O}(1).}
Since we are working in the large $L$ limit, we have proven only that the
leading term, proportional to $L$, vanishes. This, of course, is required in a
supersymmetric theory---\wit\ gives the number of bosonic ground states minus
the number of fermionic ones, because the others pair up and cancel. (For the
$k$ theory, $\tr(-1)^F=k+1$.)  Thus this calculation gives no new information,
but does provide a check on our TBA system, as well as giving as simple
example of the application of the excited-state TBA analysis \rPF\ to models
with pseudoparticles.

\subsec{$N$=1 minimal models with $\Phi _{(1,3)}$ superfield perturbation}

The S-matrix for the scattering theory coming from the $N$=1 minimal model
$SU(2)_2\otimes SU(2)_k/SU(2)_{k+2}$ perturbed in the least relevant,
supersymmetry preserving operator (the $\Phi _{(1,3)}$ superfield) was
conjectured in \refs{\rSch,\rABL}\ to be
\eqn\blsprod{S^{N=1}_k(\t )=S^{N=0}_k (\t )\otimes S^{N=1}_{k=1}(\t ),}
where $S^{N=1}_{k=1}$=$S^{N=0}_{k=2}$, the S-matrix for the perturbed
tricritical Ising model. This was conjectured because it is the simplest
S-matrix consistent with the appropriate quantum-group symmetry and the
supersymmetry \rABL.  The first part of this tensor product arises because the
Landau-Ginzburg picture \rVW\ predicts solitons interpolating between the
minima of a Chebyshev superpotential, just as in the $N$=2 case.  The second
part gives the model its supersymmetric structure, which acts nonlocally on
solitons \RSOS. The origin of this soliton structure is unclear. Presumably
there is a bosonization of this series analogous to the sine-Gordon
description of the $N$=2 $X^3-X$ model, and the solitons interpolate between
vacua of this field.

To find the TBA system, we must diagonalize the transfer matrix \Tab\
associated with the S-matrix \blsprod. As in the $N$=2 series, the S-matrix is
a tensor product, and we diagonalize the two parts separately. Each piece is
an $N$=0 minimal model, so this diagonalization has already been done \RSOSTBA,
and is discussed at the beginning of this section. As in the $N$=2 series, the
resulting two sets of pseudoparticles are coupled through the massive node:
\bigskip
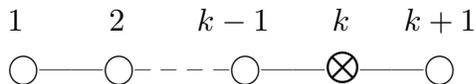
\centerline{\hbox{\rlap{\raise15pt\hbox{$\ 1\hskip.95cm 2\hskip.8cm
k-1\hskip.7cm k\hskip.6cm k+1$}}
$\bigcirc$------$\bigcirc$-- -- --
--$\bigcirc$------{\raise1pt\hbox{$\bigotimes$}}------$\bigcirc$}}
\bigskip
\noindent
It is simple to take the UV limit of this system using the analysis of the
$N$=0 minimal models. Since we now have $k+1$ nodes instead of $k$, the $x_a$
are given by \anxa\ with the substitution $k\rightarrow (k+1)$. When the
massive node is removed, we obtain a diagram of $k-1$ nodes along with a
single disconnected node. Thus the $y_a$ for $a=1\dots k$ are given by \anya,
while $y_{k+1}= 1$. Using the dilogarithm identity \amazing\ again, we find
that
\eqn\noneE{E(mR\rightarrow 0)\rightarrow
-\left({\pi\over 6R}\right)\left({3\over 2}- {12\over (k+2)(k+4)}\right),}
in agreement with the conformal results.

We note that the $k=2$ model in this series corresponds to the first model in
the $N$=2 series. However, the S-matrix \blsprod\ for $k=2$ is not identical
to the S-matrix \Smat. In fact, the particle content is different: the
particle content here is not the soliton doublet discussed above but instead
is that of the tensor product of two triple-well soliton potentials. This is
equivalent to making the vacua the nodes and the solitons the links of the
Dynkin diagram for $\widehat A_3$ or $\widehat D_4$, depending on how one does
the tensor product. However, one can recover the S-matrix elements \Smat\ from
\blsprod\ by taking an orbifold, where one mods out by a symmetry. The
orbifold procedure is a generalization of the Kramers-Wannier high- to
low-temperature duality.  This redefines the states as linear combinations of
the previous states, hence changing the S-matrix.  The new S-matrix can be
found in the manner of lattice-model orbifolds \rFG, and one recovers the
S-matrix \Smat.  This was done in \RSOS\ to relate an $N$=0 $D$-series
S-matrix (the tricritical three-state Potts model) to one in the $A$-series.
In fact, this can be done for every other model in the $D$-series, as in the
lattice models of \rFG. The orbifolding should not change the ground-state
energy, but will affect excited-state energies \rKMv.

\subsec{$SU(2)_l\otimes SU(2)_k/SU(2)_{k+l}$}

The general $SU(2)$ coset case is the obvious generalization. The S-matrix
here is \rABL
\eqn\blsprodii{S^{N=0}_{k,l}(\t )=S^{N=0}_{k} (\t )\otimes S^{N=0}_{l}(\t ),}
It follows that the TBA system is described by the diagram\

\bigskip
\centerline{\hbox{\rlap{\raise15pt\hbox{$\ 1\hskip.95cm 2\hskip.8cm
k-1\hskip.7cm k\hskip.6cm k+1\hskip.4cm k+l-1$}} $\bigcirc$------$\bigcirc$--
-- -- --$\bigcirc$------{\raise1.5pt\hbox{$\bigotimes$}}------$\bigcirc$-- --
-- --$\bigcirc$}}
\bigskip
\noindent
The $x_a$ are given by \anxa\ with the substitution $k\rightarrow k+l-1$. The
$y_a$ for $a=1,\dots,k$ are given by \anya, while for $a=k+1,\dots, k+l-1$
they are given by \anya\ with $a\rightarrow a-k$ and $k\rightarrow l$. The
result for the energy is in agreement with the central charge of the conformal
theory $c=3({k\over k+2}+{l\over l+2}-{k+l\over k+l+2})$.

\newsec{Zamolodchikov's c-function and more}

The ground state energy $E(R)$ described in the previous section in terms of
the solutions to our set of coupled integral equations can be thought of as a
Zamolodchikov `c-function' \rcfun\ for our massive theory: $E(R)=-\pi
c(mR)/6R$.  It is not obvious from the TBA equations that this
`c-function' monotonically decreases in the infra-red, though this is observed
to be the case in the numerical and analytic analysis of all physical  TBA
systems checked\foot{T.R. Klassen and E. Melzer, private communication}.

In \rCV\ another `c-function' was obtained for the
Chebyshev theories using the $N$=2 structure.  It is interesting to compare
these results.  The
c-function obtained in \rCV\ is quite simple:
\eqn\cvcfunct{c_k(mR=z)={3\over 2}z{d\over dz}u(z,r=2k/(k+2))}
where $u(z,r)$ is a solution of the differential equation
$u_{zz}+z^{-1}u_{z}=\sinh u$ (which can be transformed to Painleve III).  The
boundary conditions of $u$ are specified by (for $r<2$)
\eqn\bc{u(z,r)\sim rlogz +\ const.\ + \dots \qquad \hbox{as}
\quad z\rightarrow 0.}
The values for $r$ in the solutions \cvcfunct\ are obtained from regularity
conditions.  The only $k$ dependence is in the boundary conditions of $u$.  It
is easily seen from \bc\ that \cvcfunct\ yields the correct central charges in
the UV limit $z\rightarrow 0$.  It would be interesting to directly calculate
the next UV corrections to the `c-function' from perturbation theory around
the conformal theory and compare with that predicted by the TBA approach and
that predicted by \cvcfunct.  Comparing the IR limits ($z\rightarrow \infty$),
it can be seen that the two approaches give different results: the c-function
\cvcfunct\ has contributions corresponding to only odd numbers of particles,
whereas the TBA approach includes contributions from any number of particles.
It remains to be seen if one can relate these approaches by some projection in
the manner of \rPF. If this were possible it would be quite interesting: we
could work with an differential equation (and a well-studied one at that)
rather than the TBA system of coupled integral equations \TBA.

\newsec{Conclusions}

We have found the exact S-matrix for the $N$=2 discrete series perturbed by
the least-relevant supersymmetry-preserving operator. The nice thing about the
$N$=2 non-renormalization theorems is that they leave little guesswork; here
one does not have to conjecture the particle content.  Since our results fit
in so nicely with the $N$=0 discrete series, we believe that this lends more
support to the conjectured $N$=0 soliton picture, and that much intuition can
be gained by using the $N$=2 models to study the $N$=0 models.

Another interesting aspect is the appearance of solitons with fractional
fermion number: $\pm\half$ here, and any rational number in our next paper.
There are only a few other integrable models with such structure, and those
have only fractional charge $\pm\half$ \rSW. Therefore, this analysis may be
valuable for studying experimentally-realizable systems of fractional fermion
number. It is probably not possible to measure the S-matrix or the Casimir
energy directly, but one knows many other things about these $N$=2 systems
which may be observable. For example, all the critical exponents of these
systems are known exactly.

\bigskip\bigskip

\centerline{\bf Acknowledgements}

We are very grateful to N.\ Reshetikhin for many helpful conversations on the
TBA, and for providing the presentation of sect.\ 5.2.  We would also like to
thank C.\ Vafa for valuable discussions, and T.\ Klassen and E.\ Melzer for
their comments on the manuscript.  K.I.\ was supported by NSF grant
PHY-87-14654 and an NSF graduate fellowship, and P.F.\ was supported by DOE
grant DEAC02-89ER-40509.

\appendix{A}{Algebraic Bethe Ansatz for the 6-vertex Model}

Consider a scattering theory consisting of two species $u$ and $d$ with the
two particle S-matrix:
\eqn\abcii{\bmx{&du&ud\cr
ud&b&{\tilde c}\cr du & c &{\tilde b}\cr}\qquad \qquad \bmx{&uu &dd\cr
uu&a&0\cr dd &0&{\tilde a}\cr}.}  Our S-matrix elements are taken to satisfy
the
condition
\eqn\ff{a(\t ){\tilde a}(\t )+b(\t ){\tilde b}(\t )-c(\t )\tilde c(\t )=0.}
Drawing a scattering vertex with right or up pointing arrows to denote species
$u$ and left or down arrows to denote species $d$, the scattering theory
\abcii\ corresponds to the 6-vertex model.  Condition \ff\ is known as the
free fermion condition in the six-vertex model\rBax.  We will discuss the
algebraic Bethe ansatz technique for finding the eigenvalues of the transfer
matrix for theory \abcii .  Note that our basic $N$=2 scattering theory is of
this type; we are here being more general with future applications in mind.

The matrix $T_{ab}(\t )$ discussed in sect. 4.2 will be denoted by
\eqn\ABCD{T_{ab}(\t )=\pmatrix{A(\t )&B(\t )\cr C(\t )&D(\t )};}
$C(\t )$, for example, is the transfer matrix for bringing a particle of type
$d$ and rapidity $\t$ through $N$ particles with rapidities $\t _i$ and ending
up with a particle of type $u$ coming out.  As discussed in sect. 4.2, we will
be interested in finding the eigenvalues of the $2^N\times 2^N$ matrix $tr _o
T=A(\t )+D(\t )$.

Define $\O\equiv |\prod _{i=1}^Nd(\t _i)\>$.  Notice from \abcii\ that for any
rapidities we have $C(\t )\O =0$.  The
state $\O$ is also an eigenstate of the matrices $A$ and $D$:
\eqn\ADonO{A(\t )\O =\prod _{i=1}^Nb(\t -\t _i)\O , \quad D(\t )\O =\prod
_{i=1}^N {\tilde a}(\t -\t _i)\O.}
The algebraic Bethe ansatz is that all eigenvectors of the matrix $A +D$ can
be written as
\eqn\evects{\psi =\prod _{r=1}^nB(y_r)\O ,}
for some choices of the $y_r$ and $n$ (it follows from \abcii\ that the vector
\evects\ has $n$ $u$ particles and $N-n$ $d$ particles).  As a result of the
Yang-Baxter relation, $T_{ab}(\tht)$ satisfies the relation
\eqn\STT{S_{ac}^{a'c'}(\tht-y_r)T_{a'b}(\tht)T_{c'd}(y_r)=T_{cd'}(y_r)
T_{ab'}(\tht)S_{b'd'}^{db}(\tht-y_r)}
This yields the following
useful commutation relations for the transfer matrices:
\eqn\stt{\eqalign{A(\t )B(y_r)&={a(\t -
y_r )\over b(\t - y_r)}B(y_r)A(\t )-{c(\t -y_r)\over
b(\t -y_r)}B(\t )A(y_r)\cr
D(\t )B(y_r)&=-{a(\t -y_r)\over b(\t -y_r
)}B(y_r)A(\t )+{c(\t -y_r)\over
b(\t -y_r)}B(\t )D(y_r),\cr}}
provided $b(\t -y_r)\neq 0$. We have used the condition \ff\ to simplify
the equations.
It follows from these relations and \ADonO\ that $A(\t )+D(\t
)$ act on the states \evects\ as
\eqn\evals{[A(\t )+D(\t )]\psi= \prod _{r=1}^n{a(\t -
y_r)\over b(\t -y_r)}(\prod _{i=1}^Nb(\t -\t _i)+(-1)^n\prod
_{i=1}^N{\tilde a}(\t -\t _i))\psi  + \hbox{bad terms},}
where the `bad terms' are vectors related to \evects\ by some
$B(y_r)\rightarrow B(\t _p)$.  All such bad terms are proportional to $[
A(y_r)+(-1)^nD(y_r)]\O$ and thus vanish if
\eqn\bad{\prod _{i=1}^N{b(y_r-\t _i)\over {\tilde a}(
y_r-\t _i)}=(-1)^{n+1},}
this determines the $y_r$.  Actually, since our S-matrices all have $b(0)=0$,
the argument (using \stt ) that all bad terms are proportional to \bad\ breaks
down if some of the $y_r$ are the same.  We thus require the solutions  of
\bad\ to have no repeating values of $y_r$.  Our desired eigenvalues of $\tr
_oT (\t )$ are
thus
given by
\eqn\evs{\l (\t ;y)=\prod _{r=1}^n {a(\t -y_r)\over
b(\t -y_r)}(\prod _{i=1}^Nb(\t -\t _i)+(-1)^n\prod _{i=1}^N{\tilde a}(\t -\t
_i)),}
for any choice of distinct solutions $y_r$ and $n$ of \bad .

\listrefs

\bye